# SOLUTION OF THE NONLINEAR EQUATION OF A DIVERGENT TYPE IN THE CORNER POINT DOMAIN

## Perepelkin E. E., Sadovnikov B. I., Inozemtseva N. G.

**Abstract**
   The algorithm for generation of exact solutions of the nonlinear equation in partial derivatives of a divergent type which is included in the formulation of magnetostatics, hydro-and aerodynamics, quantum mechanics (stationary Schrödinger equation) has been suggested. The properties of smoothness of solutions in domains with corner points (piecewise smooth boundary) have been considered. The solutions with unlimited derivatives in the corner point domain have been presented on the basis of a new class of special functions.

**Introduction**
   The nonlinear equation of a divergent type is met in many mathematical and theoretical physics problems

$$\mathrm{div}\left[\mu(|\nabla u|)\nabla u\right]=0, \tag{1}$$

where $\mu$ – is a nonlinear function. The magnetostatics problem is one such examples. In the magnetostatics boundary value problem equation (1) describes the scalar potential of the magnetic field in the ferromagnet, and the function $\mu$ plays the role of the magnetic permeability of a ferromagnet [1,2]. The stationary continuity equation is another example, and it is included in the statement of the hydro-gas dynamics problem, function $\mu$ in this case plays the role of charge/mass density, and the vector field $\nabla u$ specifies the velocity flow. As it was shown in [3], the stationary continuity equation (1) leads to the stationary Schrödinger equation, where function $\mu$ corresponds to the density of the probability distribution, expressed via the wave function $\mu = \Psi\bar{\Psi} = |\Psi|^2$, and the field $\nabla u$ defines a vector field of flow probabilities.

   The majority of nonlinear boundary value problems solutions are usually calculated by numerical methods. However, if the domain in which we solve the boundary value problem is nonsmooth, there is a question concerning existence and uniqueness of the solution and smoothness of solutions in the corner point domain. Fig. 1 shows the domain with two corner points $P$ and $Q$ with apex angles $\omega_P$ and $\omega_Q$, respectively.

   As there is can be a significant error in the domain of the corner points while using the numerical solution, the knowledge of the solution behaviour is required to construct a numerical algorithm with the necessary accuracy.

   The question of the existence of particular solutions of linear differential equations of an elliptic type in domains with corners has been considered in many works.

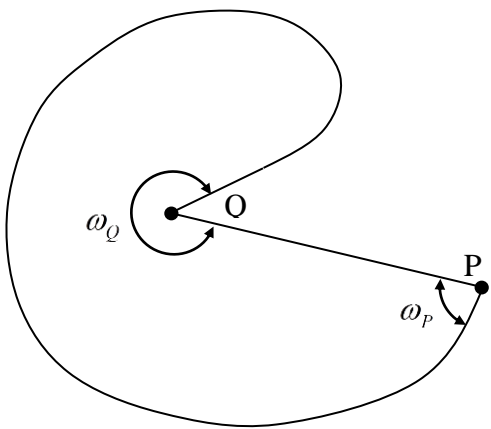

Figure 1. Corner points P and Q

   In the work of V. V. Fufaev [4] the Dirichlet problem in the planar domain with piecewise-smooth boundary for the Laplace equation has been considered. E. A. Volkov [5] investigated the conditions under which the solution of the first, second and mixed boundary-value problem for the Poisson equation in a rectangular domain $D$ belong to the $C_{k,\lambda}(D)$ class. $C_{k,\lambda}(D)$ class is regarded as the class of continuously differentiable $k$ times in $D$ functions and all $k$ derivatives satisfy the Hölder condition in $D$ with $\lambda$. In papers [7-9] he has also described various techniques for constructing different schemes for the Laplace and the Poisson equations, and suggested the method of the mesh condensation in the corner point domain, giving the same order of convergence as for regular schemes in the region with a smooth boundary.

The general case of the boundary value problem for an elliptic equation was discussed by G. I. Eskin [6] under the assumption that boundary conditions and the right part of the equation belong to the space $C^N$ (where $N$ is «sufficiently» large). V. A. Kondratiev [10] considered the case when the right part of the equation and boundary conditions belong to the space $W_2^k$ (the space of S. L. Sobolev [20-22]), and the equation coefficients are infinitely differentiable functions. In particularly, V. A. Kondratiev has shown the following

$$\sum_{i,j=1}^{2} a_{ij}(x) u_{x_i x_j} + \sum_{i=1}^{2} a_i(x) u_{x_i} + cu = f, \quad x \in D,$$

where $a_{ij}(0) = \delta_{ij}$, and the corner point is at the origin, the asymptotic behavior of the solution is $u(x) \in W_2^1(D)$ provided $f \in W_2^k(D)$ and $u|_\Gamma = \phi_0 \in W_2^{k+\frac{3}{2}}(\Gamma)$ has the following form

$$u = \sum_{0 < \frac{\pi m}{\omega_0} < k+1} \alpha_{mq} r^{\frac{\pi m}{\omega_0}} P_{mq}(r \ln^q r) + \sum_{0 \leq i_1 + i_2 \leq k+1} a_{i_1 i_2} x_1^{i_1} x_2^{i_2} +$$
$$+ \sum_{2 \leq j \leq k+1} r^{j_1} \ln^{j_2} r \theta_{j_1 j_2}(\varphi) + w,$$

where $m, j_1, j_2$ are integers; $w \in \overset{\circ}{W}_0^{k+2}(D)$; $P_{mq}$ – a polynomial the coefficients of which are infinitely differentiable functions that are linear combinations of trigonometric functions; $\theta_{j_1 j_2}$ are infinitely differentiable functions; $\omega_0$ – the value of the angle at the corner point (here it is assumed that the boundary region in the domain of the corner point is formed by straight lines). Norms $W_2^{k-\frac{1}{2}}(\Gamma)$ and $\overset{\circ}{W}_\lambda^k(D)$ are defined as

$$\|\phi\|_{W_2^{k-\frac{1}{2}}(\Gamma)}^2 = \inf \|v\|_{W_2^k(D)}^2,$$

$$\|u\|_{\overset{\circ}{W}_\lambda^k(D)}^2 = \sum_{m=0}^{k} \iint_D r^{\lambda - 2(k-m)} \left|\frac{\partial^m u}{\partial x^m}\right|^2 dx,$$

where inf is taken over all functions $v|_\Gamma = \phi$.

L. A. Oganesyan and L. A. Rukhovets [11], and of E. A. Volkov [9] in their works presented the construction of the variational-difference schemes (VRS), that use the idea of the mesh condensation in the domain of the corner point and method of variables replacement, at which the function of features is «smooth enough» in the new variables. A similar idea is presented by J. Babuska [12] for construction of VRS for the third boundary value task.

J. Babuska, M. B. Rozenzwieg [13] have described the VRS construction by the Galerkin method where test functions are taken as the product of local functions on the weight functions the corner points $r^{2\alpha}$, where $r$ is the distance to the corner point, and $\alpha \in [0,1)$. The convergence of this method is set in accordance with a weight norm.

A method of the function of features allocation, which in the domain of the corner point has the following form, is rather common

$$\psi_i(r,\varphi) = r^{\lambda_i} \ln^{p_i} r \Phi_i(\varphi), \qquad (2)$$

where $\Phi(\varphi)$ is an analytical function; $\lambda_i$ is a positive constant; $p_i$ − nonnegative integers; $(r,\varphi)$ − polar coordinate system with the center at the corner point.

In the works, for instance, of G. Fix [14], L. A. Oganesyan, L. A. Rukhovets, V. Y. Rivkind [15, 16] there is a description of a method for the VRS construction, when in order to increase the accuracy of the basic functions it is necessary to add a function of the form (2), that takes the singularity at the corner point into account. In the works of A. A. Samarskii, I. V. Fryazinov [17, 18] a similar method is used in the construction of differential schemes for various types of boundary value problems.

In the works of O. A. Ladyzhenskaya, N. N. Uraltseva [19] there is a detailed consideration ofail quasilinear equation of an elliptic type. In particular, the case of quasilinear equations with a divergent principal part, which is similar to the equation $\operatorname{div}\left[\mu(|\nabla u|)\nabla u\right]=0$, included in the formulation of magnetostatics, has been considered. There are also theorems concerning existence, uniqueness of solution $u(p)$ and limitedness of $\max|\nabla u|$ for boundary value problem in domains with a smooth boundary. In the work, for example, of L. A. Oganesyan, L. A. Rukhovets, V. J. Rivkind [15,16] for the boundary value problem with a quasilinear equation there the VRS construction in the domain with a smooth boundary is given.

However, this issue has been little studied for nonlinear equations. The difficulty is in constructing general solutions of nonlinear equation in partial derivatives. In the best case it is possible to find some particular solutions [23-26]. For example, the application of a group analysis of differential equations, that was discovered by a Norwegian mathematician Sophus Lie [27,28], using the Painlevé analysis [29-37], the construction of the Lax pairs [34,38-40], using the method of inverse scattering problem [32,34,38,41-44].

Transformations play an important role in the study of nonlinear partial differential equations. For example, the Cole-Hopf transformation allows one to reduce Burgers' equation to linear heat equation. The Korteweg-de Vries equation with the help of the Miura transformation can be represented in the form of the Lax pair. Generalization of the Miura transformation by K. Gardner made it possible to find an infinite number of conservation laws for the Korteweg-de Vries equation. These transformations allow finding the solutions of other equations if there is a known solution of one of equations, and they are representations of solutions of one equation to another solution.

In [45] we proposed the use of nonlinear transformations of Legendre [46] during the search of solutions of equation (1). As a result, the nonlinear equation (1) was reduced to a linear equation the properties of which we could explore by a well-developed apparatus of the theory of linear equations. After the solution of the linear equation is built, it can be transformed back using the inverse Legendre transformation and we can obtain the solution of the nonlinear equation (1).

In this paper there is a further extension of the approach associated with the Legendre transformation basing on the construction of exact solutions of equation (1) for certain types of nonlinear function $\mu$ which is commonly found in problems of magnetostatics.

In §1 there is a consideration of the construction of the general form of solution in the domain of the corner points using the Legendre transformation. The conditions under which the solutions of equations (1) exist, with unlimitedly growing derivatives in the domain of the corner points are presented.

In §2 for functions of the magnetic permeability $\mu$ are, smooth solutions based on the Jacobi polynomials with their subsequent mapping via nonlinear Legendre transformation, are constructed.

In §3 there are the solutions with unlimitedly growing derivatives in the domain of the corner point. While constructing such solutions, a new class of special functions represented in the form of an expansion into a series in the domain of a singular point (corner point) has been obtained.

In §4 the boundary value problem will be considered for equation (1) in the domain with corner. An increase in $|\nabla u|$ in the region around the corner point will be assessed, and an algorithm for numerical calculation of this problem will be presented.

### § 1 Statement of the problem.

As it was shown in [45] using the nonlinear Legendre transformation [46], equation

$$\operatorname{div}\left[\mu(|\nabla u|)\nabla u\right]=0 \tag{1}$$

can be reduced to a linear differential equation of the second order. The Legendre transformation is based on the fact that some surface $u(x, y)$ is defined not as a multitude of points $(u, x, y)$ but as a multitude of tangent planes. This mapping differs from a simple coordinate transformation as it matches not a point to a point, but the surface element $(x, y, u, u_x, u_y)$ to the surface element $(\xi, \eta, \omega, \omega_\xi, \omega_\eta)$. The following relations are true:

$$\begin{aligned}
&\omega(\xi,\eta)+u(x,y)=x\xi+y\eta, \\
&\xi=u_x, \quad \eta=u_y, \quad x=\omega_\xi, \quad y=\omega_\eta, \\
&u_{xx}=J\omega_{\eta\eta}, \quad u_{xy}=-J\omega_{\xi\eta}, \quad u_{yy}=J\omega_{\xi\xi},
\end{aligned} \tag{2}$$

where $J=u_{xx}u_{yy}-u_{xy}^2$ is the jacobian of the Legendre mapping. As a result, the equation (1) takes the form

$$\left[1+\xi^2 f(r)\right]\omega_{\eta\eta}-2\xi\eta f(r)\omega_{\xi\eta}+\left[1+\eta^2 f(r)\right]\omega_{\xi\xi}=0, \tag{3}$$

where $f(r)\stackrel{\text{det}}{=}\dfrac{\mu'(r)}{\mu(r)r}$, $r=\sqrt{\xi^2+\eta^2}$. Note that equation (3) is linear, unlike the original equation (1). If the solution of equation (3) satisfies the condition:

$$|\nabla\omega(\xi,\eta)|\to 0 \quad for \quad (\xi,\eta)\to\infty \tag{4}$$

then the solutions as (4) under the transformation (2) will give the solution of equation (1) with the property:

$$|\nabla u(x,y)| \to \infty \quad for \quad (x,y) \to 0. \tag{5}$$

In addition to the solutions of type (5), equation (1) can have solutions with limited $|\nabla u|$; for that the domain that is searched for the solution of the equation (3) should be limited.

During the transition to the polar coordinate system equation (3) takes the form:

$$v_{rr} + a(r)\left(\frac{1}{r}v_r + \frac{1}{r^2}v_{\varphi\varphi}\right) = 0, \tag{6}$$

where $a(r) \stackrel{det}{=} 1 + r^2 f(r)$, $\omega(\xi,\eta) = \omega(r\cos\varphi, r\sin\varphi) \stackrel{det}{=} v(r,\varphi)$. Looking for the solution of equation (6) in a factored form $v \sim R(r)\Phi(\varphi)$, we obtain two equations for the radial and angular component, respectively:

$$R'' + a(r)\left(\frac{1}{r}R' - \frac{\gamma^2}{r^2}R\right) = 0,$$
$$\Phi'' + \gamma^2 \Phi = 0, \tag{7}$$

where $\gamma = const$. Function $\Phi(\varphi)$ is represented as a superposition of functions $\sin(\gamma\varphi)$ and $\cos(\gamma\varphi)$. Function $R(r)$ is representable in the form of an expansion in a series in the domain of a singular point:

$$R_1(r) = r^\nu \sum_{k=0}^{+\infty} a_k r^k,$$
$$R_2(r) = r^{-\nu} \sum_{k=0}^{+\infty} b_k r^k \quad or \quad R_2(r) = \alpha R_1(r)\ln r + \beta r^\nu \sum_{k=0}^{+\infty} c_k r^k. \tag{8}$$

On the basis of (7), (8) the general solution of equation (6) (or (3)) can be represented in the form of a formal series

$$\omega(\xi,\eta) = v(r,\varphi) = \sum_\gamma R_\gamma(r)\Phi_\gamma(\varphi). \tag{9}$$

Then, using the Legendre transformation (2), solution (9) is to be transferred to the solution $u(x,y)$ of the nonlinear equation (1). Some particular solutions of equation (1) can be obtained in an explicit form, and their properties can be investigated. For example, if $\gamma = 0$, in accordance with (7)-(9) and (2), solution of equation (1) takes the form

$$u(x,y) = C_1 \arctg\left(\frac{y}{x}\right) + C_2 = C_1\tilde\varphi + C_2, \tag{10}$$

where $x = \tilde{r}\cos\tilde{\varphi}$, $y = \tilde{r}\sin\tilde{\varphi}$. It is seen that at the corner point (the origin) solution (10) is not defined, and value $|\nabla u| = \dfrac{C_1}{\tilde{r}}$ unlimitedly grows when approaching the origin.

Solutions corresponding to other values $\gamma$ can be used to solve boundary value problems in domains with corners. For example, the Dirichlet problem for the nonlinear equation (1)

$$\begin{cases} \operatorname{div}\left[\mu(|\nabla u|)\nabla u(p)\right] = 0, & p \in \Omega, \\ u|_\Gamma = u_0, \end{cases} \qquad (11)$$

where domain $\Omega$ is presented in Fig. 1.

## § 2 Solutions with limited derivatives

The specific form of coefficients $a_k$, $b_k$ or $c_k$ in expression (8) from §1 depends on the type of the function. In [47] with large fields for ferromagnets, the asymptotics of the function of the magnetic permeability is specified:

$$\mu(H) = 1 + \frac{M_0}{H} - \frac{M_1}{H^2}, \quad M_0 > 0, \; M_1 \geq 0. \qquad (1)$$

Fig. 2 presents the behavior of the magnetization $M$ depending on the field $H$ for the substance with the magnetic permeability (1). Were $M_0 = 0.1$, $M_1 = 0.01$ presented as examples. Note that from a physical point of view value $M_0$ sets the maximum value of magnetization, as $M(H) = \chi(H)H = (\mu(H) - 1)H$, where $\chi(H)$ is the magnetic susceptibility.

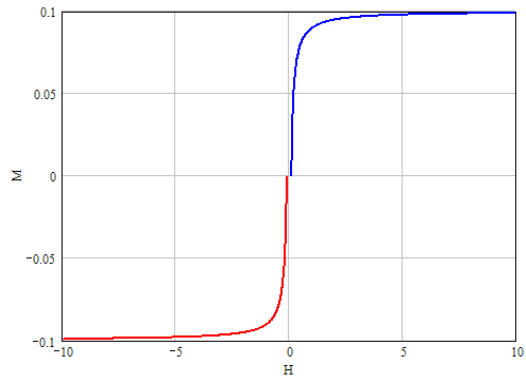

Figure 2. Dependence of magnetization $M$
from the magnetic field $H$ strength

### 2.1 Case $M_1 = 0$

Let us find function $R(r)$ with the function of the magnetic permeability (1). We start with the case when $M_1 = 0$, then

$$a(r) = \frac{r}{r + M_0}. \qquad (2)$$

Equation (7) from §1 for function $R(r)$ takes the following form:

$$R'' + \frac{1}{r+M_0}R' - \frac{\gamma^2}{r(r+M_0)}R = 0. \qquad (3)$$

The obtained equation (3) can be reduced to the Jacobi type equation on a finite interval:

$$\sigma(r)u'' + \tau(r)u' + \lambda u = 0 \qquad (4)$$

where

$$\sigma(r) = Ar^2 + Br + C, \ \tau(r) = Dr + E, \ \lambda = Const.$$

Equation (3) is a special case of equation (4) when $A=1$, $B=M_0$, $C=0$, $D=1$, $E=0$, $\lambda = -\gamma^2$ that is:

$$\sigma(r) = r^2 + M_0 r, \ \tau(r) = r, \ \lambda = -\gamma^2. \qquad (5)$$

As it is known the Jacobi equation (4) can be rewritten in the form:

$$\frac{d}{dr}\rho(r)\sigma(r)\frac{du}{dr} + \lambda\rho u = 0, \qquad (6)$$

where function $\rho(r)$ satisfies the differential ratio

$$\frac{[\rho(r)\sigma(r)]'}{\rho(r)} = \tau(r) \quad \text{or} \quad \frac{d}{dr}\ln[\rho(r)\sigma(r)] = \frac{\tau(r)}{\sigma(r)}. \qquad (7)$$

Thus, for equation (3) function $\rho(r)$ can be found from (7) concerning (5):

$$\ln[\rho(r)\sigma(r)] = \int \frac{dr}{r+M_0} = \ln[r+M_0] + C,$$

$$\rho(r) = \frac{C_0(r+M_0)}{r(r+M_0)} = \frac{C_0}{r}. \qquad (8)$$

As the solution of equation (6) is defined up to a multiplier, then, without limiting the generality the constant value $C_0$ in (8) can be equal to one. As it is known the solution of the Jacobi equation are the polynomials that are obtained by the following formula:

$$P_n(r) = \frac{B_n}{\rho(r)}\frac{d^n}{dr^n}[\rho(r)\sigma(r)^n], \qquad (9)$$

where $B_n$ is a constant value that can be chosen from the normalization condition of polynomials on interval $[-M_0, 0]$. Let us note that polynomials (9) are orthogonal on interval $[-M_0, 0]$.

Let us write the first three polynomials (9):

$$P_1(r) = B_1 r,$$
$$P_2(r) = B_2\left(6r^2 + 4M_0 r\right), \quad (10)$$
$$P_3(r) = 6B_3\left(10r^3 + 12M_0 r^2 + 3M_0^2 r\right).$$

Graphs of polynomials (10) are presented in Fig.3 with $M_0 = 1$.

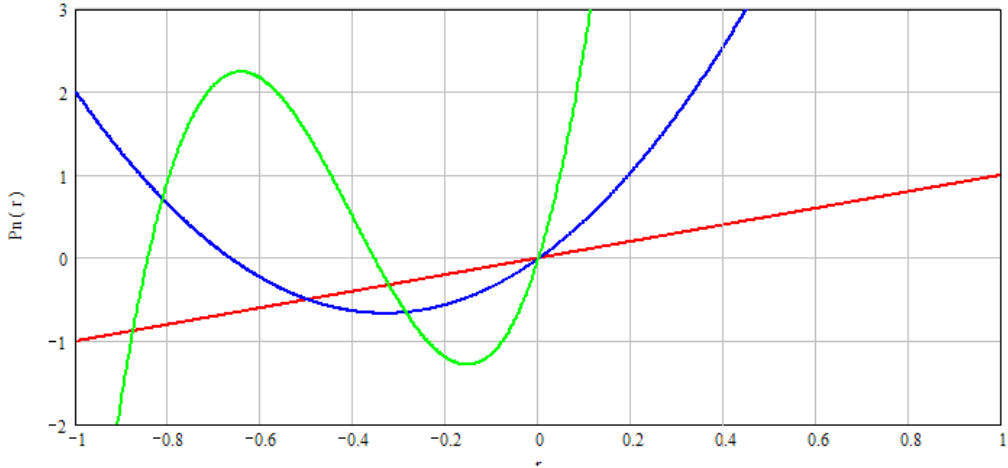

Figure 3. Polynomials graphs: $P_1(r)$ − red, $P_2(r)$ − blue, $P_3(r)$ − green

Note that, although the polynomials are orthogonal on interval $[-M_0, 0]$, the most interesting polynomials are on interval $[0, +\infty)$ from a physical point of view, as the radius $r$ cannot be negative.

Taking relation (5) into account one can find the eigenvalues $\lambda$ in equation (6) by the known formula:

$$\lambda_n = -n\tau' - \frac{n(n-1)}{2}\sigma'' = -n - n(n-1) = -n^2 = -\gamma^2,$$
$$\gamma_n = n. \quad (11)$$

Formally, the solution of equation (6) from §1 can be represented by the following sequence:

$$v(r, \varphi) = \sum_n P_n(r)\left(a_n \cos(n\varphi) + b_n \sin(n\varphi)\right). \quad (12)$$

To obtain solutions $u(x, y)$ of the nonlinear equation (1) from §1 it is necessary to apply the inverse Legendre transformation to the solution of (12). For that purpose it is necessary to solve the system of equations:

$$\begin{cases} x = \omega_\xi(\xi, \eta), \\ y = \omega_\eta(\xi, \eta), \end{cases} \text{ where } \omega(r\cos\varphi, r\sin\varphi) = v(r, \varphi), \quad (13)$$

in relation to $\xi = \xi(x,y)$, $\eta = \eta(x,y)$ and use the formula:

$$u(x,y) = x \cdot \xi(x,y) + y \cdot \eta(x,y) - \omega(\xi(x,y), \eta(x,y)). \tag{14}$$

Fig. 4 shows a graph of equation (1) from §1 corresponding to solution (6) from §1:

$$v_2(r,\varphi) = P_2(r)\cos(2\varphi), \tag{15}$$

with the function of the magnetic permeability (1) with $M_1 = 0$. Fig.5 shows a graph of the solution of equation (1) from §1, corresponding to solution (6) from §1:

$$v_3(r,\varphi) = P_3(r)\cos(3\varphi). \tag{16}$$

As it is seen, the diagrams of the solutions (Fig. 4,5) have limited derivatives. However, Fig.5 shows that the solution is not continuous along the OX axis. The graphs of other solutions of equation (1) from §1 that correspond to solutions of equation (6) from §1 of the form (12) can be made analogically.

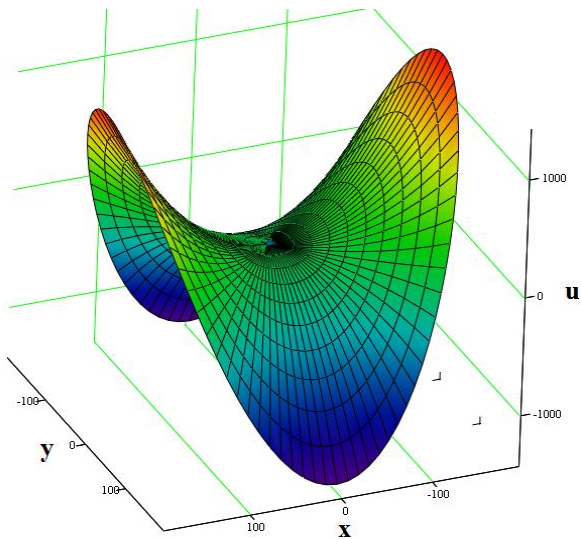
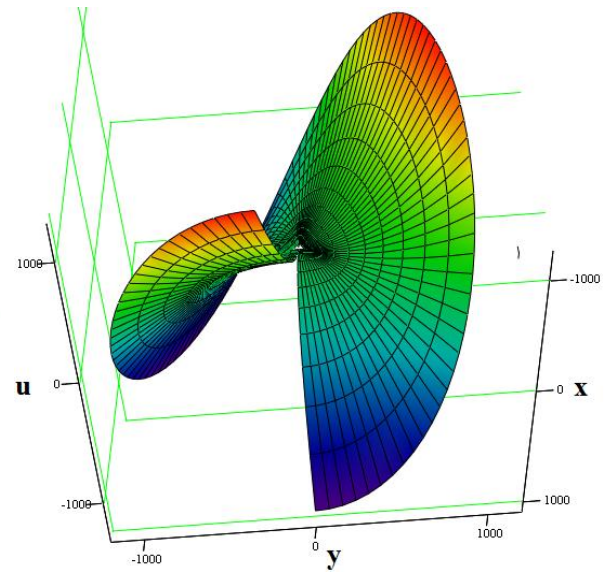

Figure 4. Solution of equation (1) from §1 corresponding to solution (15)

Figure 5. Solution of equation (1) from §1 corresponding to solution (16)

## 2.2 Case $M_1 \neq 0$

Let us consider the case when $M_1 \neq 0$, then expression (2) takes the form:

$$a(r) = \frac{r^2 + M_1}{r^2 + M_0 r - M_1}. \tag{17}$$

Note that in the particular case with $M_1 = 0$, expression (17) will be transformed into (2). Equation (7) from §1 with (17) will be:

$$R'' + \frac{r^2 + M_1}{r(r^2 + M_0 r - M_1)} R' - \frac{\gamma^2 (r^2 + M_1)}{r^2 (r^2 + M_0 r - M_1)} R = 0. \tag{18}$$

The solution will be sought in the form of a series:

$$R(r) = r^\nu \sum_{k=0}^{+\infty} a_k r^k. \tag{19}$$

Differentiating series (19) and substituting it into equation (18), we obtain:

$$r^2 (r^2 + M_0 r - M_1) \left[ \nu(\nu-1) r^{\nu-2} \sum_{k=0}^{+\infty} a_k r^k + 2\nu r^{\nu-1} \sum_{k=1}^{+\infty} a_k k r^{k-1} + r^\nu \sum_{k=2}^{+\infty} a_k k (k-1) r^{k-2} \right] +$$
$$+ r(r^2 + M_1) \left[ \nu r^{\nu-1} \sum_{k=0}^{+\infty} a_k r^k + r^\nu \sum_{k=1}^{+\infty} a_k k r^{k-1} \right] - \gamma^2 (r^2 + M_1) r^\nu \sum_{k=0}^{+\infty} a_k r^k = 0. \tag{20}$$

Equating coefficients with identical degrees by $r$, one can obtain recurrence relations on the coefficients $a_k$. For example, for $r^\nu$, we obtain:

$$-\nu(\nu-1) a_0 M_1 + \nu a_0 M_1 - \gamma^2 a_0 M_1 = 0, \tag{21}$$

As $a_0 \neq 0$, $M_1 \neq 0$, then from (21) we obtain:

$$\nu^2 - 2\nu + \gamma^2 = 0, \quad D = 4(1-\gamma^2) \geq 0, \quad \text{for } |\gamma| \leq 1,$$
$$\nu_{1,2} = 1 \pm \sqrt{1-\gamma^2}. \tag{22}$$

For the coefficients with $r^{\nu+1}$, we obtain:

$$-\nu(\nu-1) M_1 a_1 + \nu(\nu-1) M_0 a_0 - 2\nu M_1 a_1 + \nu M_1 a_1 + M_1 a_1 - \gamma^2 M_1 a_1 = 0,$$
$$-M_1 a_1 (\nu^2 + \gamma^2 - 1) + \nu(\nu-1) M_0 a_0 = 0. \tag{23}$$

Taking (22) into account, (23) will give the value of coefficient $a_1$ via $a_0$:

$$a_1 = a_0 \frac{M_0}{M_1} \frac{\nu(\nu-1)}{2\nu-1}. \tag{24}$$

Similarly, for the coefficients with $r^{\nu+2}$ taking (22) into account, we obtain:

$$a_0 \left[ \nu(\nu-1) + \nu - \gamma^2 \right] + a_1 M_0 \left[ \nu(\nu-1) + 2\nu \right] + a_2 M_1 \left[ -\nu(\nu-1) - 4\nu - 2 + \nu + 2 - \gamma^2 \right] = 0,$$
$$a_0 (\nu^2 - \gamma^2) + a_1 M_0 \nu(\nu+1) - (\nu^2 + 2\nu + \gamma^2) a_2 M_1 = 0,$$

$$a_2 = \frac{a_0(v^2 - \gamma^2) + a_1 M_0 v(v+1)}{4vM_1}. \tag{25}$$

Expression (25) for coefficient $a_2$ taking (24) into account for coefficient $a_1$ can be expressed only by the coefficient $a_0$ in the form:

$$a_2 = \frac{a_0(v^2 - \gamma^2)}{4vM_1} + \frac{a_0 M_0^2}{4vM_1^2} \frac{v^2(v^2-1)}{2v-1} = a_0 \frac{(v^2-\gamma^2)(2v-1)M_1 + M_0^2 v^2(v^2-1)}{4vM_1^2(2v-1)}. \tag{26}$$

Let us find a general form of the expression for an arbitrary number $k$ of coefficient $a_k$. Let us rewrite expression (20) in the form:

$$r^2(r^2 + M_0 r - M_1)\left[v(v-1)r^{v-2}\sum_{k=0}^{+\infty}a_k r^k + 2vr^{v-1}\sum_{k=1}^{+\infty}a_k k r^{k-1} + r^v \sum_{k=2}^{+\infty}a_k k(k-1)r^{k-2}\right] +$$

$$+ r(r^2 + M_1)\left[vr^{v-1}\sum_{k=0}^{+\infty}a_k r^k + r^v \sum_{k=1}^{+\infty}a_k k r^{k-1}\right] - \gamma^2(r^2 + M_1)r^v \sum_{k=0}^{+\infty}a_k r^k = 0,$$

$$\sum_{k=0}^{+\infty} r^k \left\{ v(v-1)a_k (r^2 + M_0 r - M_1) + 2v a_{k+1}(k+1)(r^3 + M_0 r^2 - M_1 r) + \right.$$
$$+ a_{k+2}(k+2)(k+1)(r^4 + M_0 r^3 - M_1 r^2) + v a_k (r^2 + M_1) + \tag{27}$$
$$\left. + a_{k+1}(k+1)(r^3 + M_1 r) - \gamma^2 a_k (r^2 + M_1) \right\} = 0.$$

Expression in the curly brackets (27) can be rewritten in the form:

$$r^4\left[a_{k+2}(k+2)(k+1)\right] + r^3\left[2v(k+1)a_{k+1} + a_{k+2}(k+2)(k+1)M_0 + a_{k+1}(k+1)\right] +$$
$$+ r^2\left[v(v-1)a_k + 2v a_{k+1}(k+1)M_0 - M_1 a_{k+2}(k+2)(k+1) + v a_k - \gamma^2 a_k\right] + \tag{28}$$
$$+ r\left[v(v-1)a_k M_0 - 2v a_{k+1}(k+1)M_1 + a_{k+1}(k+1)M_1\right] -$$
$$- v(v-1)a_k M_1 + v a_k M_1 - \gamma^2 a_k M_1 = 0.$$

Expression (28) should be in the curly brackets of expression (27), therefore equating the coefficients at equal degrees by $r$ in expression (27) with (28), we obtain:

$$a_{k+2}(k+2)(k+1) + 2v(k+2)a_{k+2} + a_{k+3}(k+3)(k+2)M_0 + a_{k+2}(k+2) +$$
$$+ v(v-1)a_{k+2} + 2v a_{k+3}(k+3)M_0 - M_1 a_{k+4}(k+4)(k+3) + v a_{k+2} - \gamma^2 a_{k+2} + \tag{29}$$
$$+ v(v-1)a_{k+3}M_0 - 2v a_{k+4}(k+4)M_1 + a_{k+4}(k+4)M_1 -$$
$$- v(v-1)a_{k+4}M_1 + v a_{k+4}M_1 - \gamma^2 a_{k+4}M_1 = 0.$$

Making elementary transformations in (29) with (22) we obtain the ratios between coefficients $a_{k+4}$ and $a_{k+2}, a_{k+3}$:

$$a_{k+4}M_1\left[-(k+4)(k+3)-2\nu(k+4)+(k+4)-\nu(\nu-1)+\nu-\gamma^2\right]+$$
$$+a_{k+3}M_0\left[(k+3)(k+2)+2\nu(k+3)+\nu(\nu-1)\right]+$$
$$+a_{k+2}\left[(k+2)(k+1)+2\nu(k+2)+(k+2)+\nu(\nu-1)+\nu-\gamma^2\right]=0,$$

$$-a_{k+4}M_1(k+4)(k+2+2\nu)+$$
$$+a_{k+3}M_0\left[(k+3)(k+2+2\nu)+\nu(\nu-1)\right]+$$
$$+a_{k+2}\left[(k+2)(k+2\nu+2)+\nu^2-\gamma^2\right]=0,$$

or

$$a_{k+4}=\frac{a_{k+3}M_0\left[(k+3)(k+2+2\nu)+\nu(\nu-1)\right]}{M_1(k+4)(k+2+2\nu)}+\frac{a_{k+2}\left[(k+2)(k+2\nu+2)+\nu^2-\gamma^2\right]}{M_1(k+4)(k+2+2\nu)},$$

or

$$a_{k+2}=\frac{a_{k+1}M_0\left[(k+1)(k+2\nu)+\nu(\nu-1)\right]}{M_1(k+2)(k+2\nu)}+\frac{a_k\left[k(k+2\nu)+\nu^2-\gamma^2\right]}{M_1(k+2)(k+2\nu)}. \tag{30}$$

Let us construct graphs of solutions of equation (1) from §1 and that corresponds to (19) with coefficients (30), taking conditions on the values $\nu$ of (22) into account. Fig. 6 shows a graph of the function solution corresponding to the values:

$$\gamma=0.9,\quad \nu_1=1+\sqrt{1-\gamma^2}\approx 1.436, \tag{31}$$
$$M_0=1,\quad M_1=0.01,$$
$$v(r,\varphi)=r^{\nu_1}\sum_{k=0}^{+\infty}a_k r^k \cos(\nu_1\varphi).$$

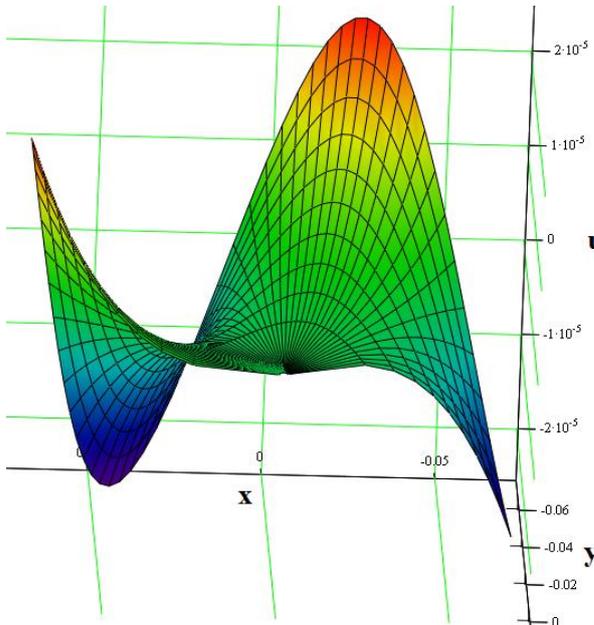

Figure 6. Solution of equation (1) from §1 corresponding to values (31)

Fig. 6 shows the solution of equation (1) from §1 at the origin as smooth and it has limited first derivatives. The limitidness of the first derivatives and continuity of the solution at the origin follows from the Legendre transformation (13)-(14), and condition $\nu_1>1$. While considering the second root $\nu_2$ of the characteristic equation (22), we obtain a value that is less than one:

$$\gamma=0.9,\quad \nu_2=1-\sqrt{1-\gamma^2}\approx 0.564. \tag{32}$$

As under the Legendre transformation (13)-(14) there is a differentiated function $v(r,\varphi)$ that is in the area of the origin $v(r,\varphi)\sim r^{\nu_2}$, where due to (32) $\nu_2<1$, then the point at which the solution of equation (1) from §1 has the limited derivatives will go to infinity. Therefore condition (31) gives a smooth solution of equation (1) from §1 at the origin, and condition (32) in the infinitely remote point.

**2.3 Remark**

Note that in the case when $M_1 = 0$, solution (19) shall transform into solutions of the form (10). Let us check this fact. Expression (23) with $M_1 = 0$ will be the following:

$$v(v-1) = 0 \Rightarrow v_1 = 0, \; v_2 = 1. \tag{33}$$

Taking (33) into account, expression (20) takes the form:

$$(r + M_0)\left[2v\sum_{k=1}^{+\infty} a_k k r^k + \sum_{k=2}^{+\infty} a_k k(k-1) r^k \right] + v\sum_{k=0}^{+\infty} a_k r^{k+1} +$$

$$+ \sum_{k=1}^{+\infty} a_k k r^{k+1} - \gamma^2 \sum_{k=0}^{+\infty} a_k r^{k+1} = 0,$$

or

$$\sum_{k=0}^{+\infty} r^k \left\{ (r^2 + M_0 r) 2v a_{k+1}(k+1) + (r^2 + M_0 r) a_{k+2}(k+2)(k+1) r + \right.$$

$$\left. + v a_k r + a_{k+1}(k+1) r^2 - \gamma^2 a_k r \right\} = 0. \tag{34}$$

Expression in curly brackets (34) can be rewritten as:

$$r^3 a_{k+2}(k+2)(k+1) + r^2 \left[ a_{k+1}(k+1) + 2v a_{k+1}(k+1) + M_0 a_{k+2}(k+2)(k+1) \right] +$$

$$+ r \left[ v a_k - \gamma^2 a_k + 2v a_{k+1}(k+1) M_0 \right]. \tag{35}$$

Substituting (35) into (34) and equating the coefficients at identical degrees by $r$, we obtain:

$$a_{k+2}(k+2)(k+1) + a_{k+2}(k+2) + 2v a_{k+2}(k+2) + M_0 a_{k+3}(k+3)(k+2) +$$

$$+ v a_{k+2} - \gamma^2 a_{k+2} + 2v a_{k+3}(k+3) M_0 = 0,$$

or

$$a_{k+2}\left[(k+2)(k+1) + (k+2) + 2v(k+2) + v - \gamma^2\right] + a_{k+3} M_0 \left[(k+3)(k+2) + 2v(k+3)\right] = 0,$$

$$a_{k+2}\left[(k+2)(k+2+2v) + v - \gamma^2\right] + a_{k+3} M_0 (k+3)(k+2+2v) = 0.$$

Making the substitution $k+2 \to k$, we obtain:

$$a_k \left[ k(k+2v) + v - \gamma^2 \right] + a_{k+1} M_0 (k+1)(k+2v) = 0,$$

$$a_{k+1} = -a_k \frac{k(k+2v) + v - \gamma^2}{M_0 (k+1)(k+2v)}. \tag{36}$$

Let us consider expression (36) for coefficients $a_k$ with different values $v_{1,2}$ (33). When $v = v_1 = 0$, expression (36) takes the form:

$$a_k \left[ k^2 - \gamma^2 \right] + a_{k+1} M_0 (k+1) k = 0,$$

$$k = 0: -\gamma^2 a_0 = 0 \Rightarrow a_0 = 0,$$

$$k = 1: a_2 = -a_1 \frac{1-\gamma^2}{2M_0},$$

$$k = 2: a_3 = -a_2 \frac{4-\gamma^2}{6M_0} = a_1 \frac{1-\gamma^2}{2M_0} \frac{4-\gamma^2}{6M_0},$$

$$k = 3: a_4 = -a_3 \frac{9-\gamma^2}{12M_0} = -a_1 \frac{1-\gamma^2}{2M_0} \frac{4-\gamma^2}{6M_0} \frac{9-\gamma^2}{12M_0},$$

...

$$k = n: a_{n+1} = -a_n \frac{n^2 - \gamma^2}{n(n+1)M_0} = (-1)^n a_1 \frac{1-\gamma^2}{2M_0} \frac{4-\gamma^2}{6M_0} \frac{9-\gamma^2}{12M_0} \cdots \frac{n^2 - \gamma^2}{n(n+1)M_0}. \tag{37}$$

If $\gamma = 1$, then from (37) it follows that $a_1 \neq 0$, $a_2 = a_3 = ... = 0$. In this case, solution (19) takes the form

$$R(r) = a_1 r \ \ for \ v = v_1 = 0 \ and \ \gamma = 1. \tag{38}$$

The obtained solution (38) coincides with solution $P_1(r) = B_1 r$ (10) with $B_1 = a_1$. For $\gamma = 2$ from (37) we obtain $a_1 \neq 0$, $a_2 = a_1 \frac{3}{2M_0}$, $a_3 = a_4 = ... = 0$. In this case, solution (19) has the form:

$$R(r) = \frac{a_1}{2M_0} \left( 3r^2 + 2M_0 r \right) \ for \ v = v_1 = 0 \ and \ \gamma = 2 \tag{39}$$

Solution of (39) coincides with the second solution from (10) $P_2(r) = B_2 \left( 6r^2 + 4M_0 r \right)$ with $B_2 = \frac{a_1}{4M_0}$. A similar convergence of solutions (19) and (10) are valid for all other integer values $\gamma$ with $v = v_1 = 0$.

Let us consider the case when $v = v_2 = 1$, then expression for the coefficients (36) takes the form:

$$a_{k+1} = -a_k \frac{k(k+2) + 1 - \gamma^2}{M_0 (k+1)(k+2)} \tag{40}$$

Expression (40) with different values $k$ gives the relationship:

$$k = 0: a_1 = -a_0 \frac{1-\gamma^2}{2M_0},$$

$$k = 1: a_2 = -a_1 \frac{4-\gamma^2}{6M_0} = a_0 \frac{1-\gamma^2}{2M_0} \frac{4-\gamma^2}{6M_0},$$

...

$$k=n: a_{n+1}=-a_n \frac{n(n+2)+1-\gamma^2}{M_0(n+1)(n+2)}=(-1)^{n+1} a_0 \frac{1-\gamma^2}{2M_0} \frac{4-\gamma^2}{6M_0} \cdots \frac{n(n+2)+1-\gamma^2}{M_0(n+1)(n+2)}. \quad (41)$$

As before, we shall consider expression (41) for different integer values $\gamma$. If $\gamma=1$ from (41) it follows that $a_0 \neq 0$, $a_1 = a_2 = ... = 0$, from (19) we obtain:

$$R(r) = a_0 r \text{ for } v = v_1 = 1 \text{ and } \gamma = 1. \quad (42)$$

Solution (42) coincides with the first solution (10) with $B_1 = a_1$. In the case where $\gamma = 2$ from (41) we obtain conditions on the coefficients $a_0 \neq 0$, $a_1 = a_0 \frac{3}{2M_0}$, $a_2 = a_3 = ... = 0$, that is, solution (19) has the form:

$$R(r) = r a_0 \left(1 + \frac{3}{2M_0} r\right) = \frac{a_0}{2M_0}\left(3r^2 + 2M_0 r\right) \text{ for } v = v_1 = 1 \text{ and } \gamma = 1. \quad (43)$$

Solution (43) coincides with the second solution of (10) with $B_2 = \frac{a_0}{4M_0}$. A similar convergence of solutions (19) and (10) are true for other integer values $\gamma$.

Thus, we can see that, with both values $v_{1,2}$ and $\gamma = n$, solution (19) with the coefficients (30) at $M_1 = 0$ transforms into solution (10)/(9).

### §3 Solutions with unlimited derivatives

#### 3.1 Case $M_1 = 0$

From the point of view of the consideration of the solutions of equation (1) from §1 with unlimited derivatives in the domain of the corner points (in this case the origin of coordinates), it is necessary to make the coordinate transformation in equation (3) from §2 of $t = 1/r$ type, and we obtain:

$$T'' + \frac{1+2M_0 t}{t(1+M_0 t)} T' - \frac{\gamma^2}{t^2(1+M_0 t)} T = 0, \quad (1)$$

where $T(t) = R(1/t)$. Let us construct the solution of equation (1) in the area of a singular point $t = 0$. The solution will be sought in the form of a series:

$$T(t) = t^v \sum_{k=0}^{+\infty} a_k t^k. \quad (2)$$

Differentiating (2) and substituting it into equation (1), we obtain:

$$\left(t^2 + M_0 t^3\right)\left[\nu(\nu-1)t^{\nu-2}\sum_{k=0}^{+\infty}a_k t^k + 2\nu t^{\nu-1}\sum_{k=1}^{+\infty}a_k k t^{k-1} + t^\nu \sum_{k=2}^{+\infty}a_k k(k-1)t^{k-2}\right] +$$

$$+\left(t + 2M_0 t^2\right)\left[\nu t^{\nu-1}\sum_{k=0}^{+\infty}a_k t^k + t^\nu \sum_{k=1}^{+\infty}a_k k t^{k-1}\right] - \gamma^2 t^\nu \sum_{k=0}^{+\infty} a_k t^k = 0. \qquad (3)$$

Equating the coefficients at identical degrees by $t$, one can obtain recurrent relations on coefficients $a_k$. For example, for $t^\nu$ we obtain:

$$\nu(\nu-1)a_0 + \nu a_0 - \gamma^2 a_0 = 0, \qquad (4)$$

As $a_0 \neq 0$, from (4) we obtain:

$$\nu^2 = \gamma^2, \quad \nu = \pm|\gamma|. \qquad (5)$$

For the coefficients $t^{\nu+1}$, we obtain:

$$a_1\left[\nu^2 + 2\nu + 1 - \gamma^2\right] + a_0 M_0 \nu(\nu+1) = 0. \qquad (6)$$

Taking (5) into account, expression (6) will give the value of coefficient $a_1$ via $a_0$:

$$a_1 = -a_0 M_0 \frac{\nu(\nu+1)}{2\nu+1}. \qquad (7)$$

Similarly, for the coefficients with $t^{\nu+2}$ and taking (5) into account, we obtain:

$$a_2\left[\nu^2 + 4\nu + 4 - \gamma^2\right] + a_1 M_0\left[\nu^2 + 3\nu + 2\right] = 0,$$

$$a_2 = -a_1 M_0 \frac{\nu^2 + 3\nu + 2}{4(\nu+1)} = -a_1 M_0 \frac{(\nu+1)(\nu+2)}{4(\nu+1)} = -a_1 M_0 \frac{(\nu+2)}{4}. \qquad (8)$$

Expression (8) for coefficient $a_2$ taking (7) into account for coefficient $a_1$ can be expressed only through coefficient $a_0$ in the form of:

$$a_2 = a_0 M_0^2 \frac{\nu(\nu+1)}{2\nu+1}\frac{(\nu+2)}{4}. \qquad (9)$$

Let us find a general view of the expression for an arbitrary number $k$ coefficient $a_k$. We start with rewriting expression (3) in the form of:

$$\sum_{k=0}^{+\infty} t^k \{\nu(\nu-1)a_k + \nu(\nu-1)M_0 t + \nu a_k + 2M_0 \nu a_k t - \gamma^2 a_k + 2\nu a_{k+1}(k+1)t +$$

$$+2\nu M_0 a_{k+1}(k+1)t^2 + a_{k+1}(k+1)t + 2M_0 a_{k+1}(k+1)t^2 + a_{k+2}(k+2)(k+1)t^2 + \qquad (10)$$

$$+a_{k+2}(k+2)(k+1)M_0 t^3\} = 0.$$

The expression in curly brackets (10) can be rewritten in the form:

$$t^3 M_0 a_{k+2}(k+2)(k+1) + t^2 \left[ 2M_0 a_{k+1}(k+1) + a_{k+2}(k+2)(k+1) + \right.$$
$$\left. +2\nu M_0 a_{k+1}(k+1) \right] + t\left[ \nu(\nu-1) A a_k + 2M_0 \nu a_k + 2\nu a_{k+1}(k+1) + \right. \quad (11)$$
$$\left. + a_{k+1}(k+1) \right] + \left[ \nu(\nu-1) a_k + \nu a_k - \gamma^2 a_k \right] = 0.$$

Expression (11) shall be in curly brackets of expression (10), therefore equating the coefficients at identical degrees by $t$ in expression (10) with (11), we obtain:

$$M_0 a_{k+2}(k+2)(k+1) + 2M_0 a_{k+2}(k+2) + a_{k+3}(k+3)(k+2) +$$
$$+ 2\nu M_0 a_{k+2}(k+2) + \nu(\nu-1) M_0 a_{k+2} + 2A\nu a_{k+2} + 2\nu a_{k+3}(k+3) + \quad (12)$$
$$+ a_{k+3}(k+3) + \nu(\nu-1) a_{k+3} + \nu a_{k+3} - \gamma^2 a_{k+3} = 0.$$

Performing elementary transformations in (12), taking (5) into account, we obtain the ratios between coefficients $a_{k+3}$ and $a_{k+2}$:

$$a_{k+3}(M_0, \nu) = -M_0 a_{k+2}(M_0, \nu) \frac{(k+2)(k+2\nu+3) + \nu(\nu+1)}{(k+3)(k+2\nu+3)},$$

or

$$a_{k+3}(M_0, \nu) = -M_0 a_{k+2}(M_0, \nu) \frac{(k+\nu+2)(k+\nu+3)}{(k+3)(k+2\nu+3)}, \quad (13)$$

or

$$a_{k+2}(M_0, \nu) = -M_0 a_{k+1}(M_0, \nu) \frac{(k+\nu+1)(k+\nu+2)}{(k+2)(k+2\nu+2)}.$$

Expression (13) defines a recurrent relation between the coefficients of series (2). To obtain a general expression for coefficient $a_k$, let us write coefficients $a_3$, $a_4$, $a_5$ using equation (3), (7) and (9), we obtain:

$$a_3 = -M_0 a_2 \frac{(\nu+2)(\nu+3)}{3(2\nu+3)} = -a_0 M_0^3 \frac{\nu(\nu+1)(\nu+2)^2(\nu+3)}{3 \cdot 4(2\nu+1)(2\nu+3)} =$$
$$= -a_0 M_0^3 \frac{\nu(\nu+1)^2(\nu+2)^2(\nu+3)}{2 \cdot 3(2\nu+1)(2\nu+2)(2\nu+3)}, \quad (14)$$

$$a_4 = -M_0 a_3 \frac{(\nu+3)(\nu+4)}{4(2\nu+4)} = a_0 M_0^4 \frac{\nu(\nu+1)^2(\nu+2)^2(\nu+3)^2(\nu+4)}{2 \cdot 3 \cdot 4(2\nu+1)(2\nu+2)(2\nu+3)(2\nu+4)}, \quad (15)$$

$$a_5 = -M_0 a_4 \frac{(\nu+4)(\nu+5)}{5(2\nu+5)} = -a_0 M_0^5 \frac{\nu(\nu+1)^2(\nu+2)^2(\nu+3)^2(\nu+4)^2(\nu+5)}{2 \cdot 3 \cdot 4 \cdot 5(2\nu+1)(2\nu+2)(2\nu+3)(2\nu+4)(2\nu+5)}. \quad (16)$$

Analyzing expressions (14)-(16) we obtain the general formula for coefficient $a_k$:

$$a_k = (-1)^k a_0 M_0^k \frac{2\nu^2 (\nu+1)^2 (\nu+2)^2 (\nu+3)^2 (\nu+4)^2 \ldots (\nu+k)^2}{k! 2\nu(2\nu+1)(2\nu+2)(2\nu+3)(2\nu+4)\ldots(2\nu+k)(\nu+k)} =$$

$$= a_0 \frac{(-1)^k M_0^k 2(\nu)_{k+1}^2}{k!(2\nu)_{k+1}(\nu+k)}, \qquad (17)$$

where designation $(\nu)_{k+1} = \nu(\nu+1)\ldots(\nu+k)$ is used, and it is associated with gamma function ratio:

$$(\nu)_k = \frac{\Gamma(\nu+k)}{\Gamma(\nu)}.$$

As constant $a_0$ is arbitrary, for convenience we choose it equal to $1/2$, in this case the final expression for coefficient $a_k$ takes the form:

$$a_k(M_0, \nu) = \frac{(-1)^k M_0^k (\nu)_{k+1}^2}{k!(2\nu)_{k+1}(\nu+k)}. \qquad (18)$$

Substituting expression (18) into (2) we obtain the solution of equation (1) in the form:

$$T_{1,2}^{(M_0)}(t) = t^{\pm|\gamma|} \sum_{k=0}^{+\infty} \frac{(-1)^k M_0^k (\gamma)_{k+1}^2}{k!(2\gamma)_{k+1}(\gamma+k)} t^k. \qquad (19)$$

Note that if $|\gamma| \neq 0$, then one of solutions $T_{1,2}^{(M_0)}(t)$ in $t=0$ has zero of $|\gamma|$ order, and the second solution has a pole of the same order at the same point. In the case with $|\gamma|=0$, there is the degenerate case, which was presented earlier in [45].

### 3.1.1 Convergence

Let us consider the question of convergence of series (19). We are examining the absolute convergence using d'Alembert's characteristic with a fixed value $\gamma$.

$$\lim_{k \to +\infty} \left| \frac{u_{k+1}}{u_k} \right| = \lim_{k \to +\infty} \frac{|M_0|^{k+1} (\gamma)_{k+1}^2 (\gamma+k+1)^2 |t|(2\gamma)_{k+1}(\gamma+k)k!}{(k+1)!(2\gamma)_{k+1}(2\gamma+k+1)(\gamma+k+1)(\gamma)_{k+1}^2 |M_0|^k} =$$

$$= \lim_{k \to +\infty} \frac{|M_0||t|(\gamma+k)(\gamma+k+1)^2}{(k+1)(2\gamma+k+1)(\gamma+k+1)} = \lim_{k \to +\infty} \frac{|M_0||t|\left(\frac{\gamma}{k}+1\right)\left(1+\frac{\gamma+1}{k}\right)^2}{\left(1+\frac{1}{k}\right)\left(1+\frac{2\gamma+1}{k}\right)\left(1+\frac{\gamma+1}{k}\right)} =$$

$$= \lim_{k \to +\infty} |M_0||t| = |M_0||t| < 1.$$

Thus, the area of convergence of series (19) has the form:

$$|t| < \frac{1}{|M_0|}. \qquad (20)$$

As value $t$ is the radius vector, then condition (20) can be rewritten as:

$$0 \leq t < \frac{1}{|M_0|}. \qquad (21)$$

Thus, the maximum magnetization $M_0$ determines the radius of convergence of series (19). Fig. 7 shows graphs of solution (19) for various values $\gamma$ with zero at the point $t = 0$.

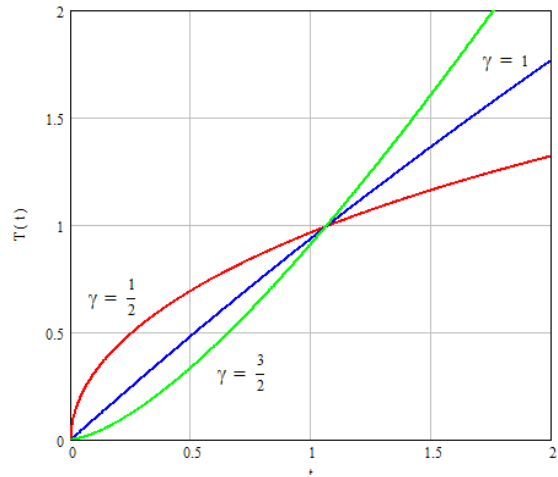

Figure 7. Function $T(t)$

Let us create the graph of a particular solution of equation (1) from §1, corresponding to the function:

$$v(r, \varphi) = T_{1,2}^{(M_0)}\left(\frac{1}{r}\right)\cos(\gamma\varphi). \qquad (22)$$

Applying the Legendre transformation (13)-(14) to function (22), we obtain particular solutions of equation (1) from §1. Fig. 8 shows the graph of the function corresponding to $\gamma = 3$ and $\gamma = 1/2$ in Fig. 9.

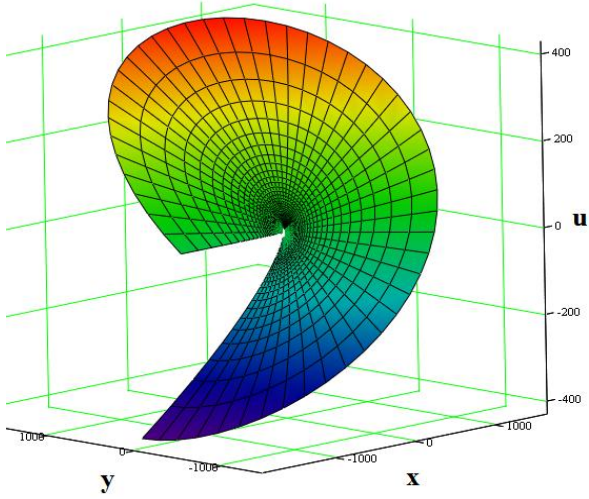 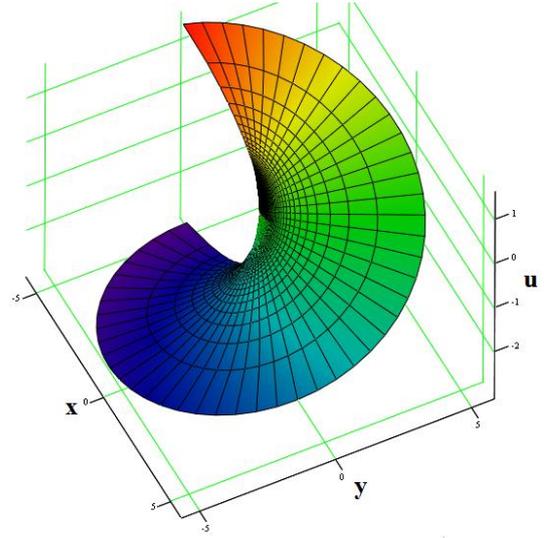

Figure 8. Solution of equation (1) from §1 with $\gamma = 3$

Figure 9. Solution of equation (1) from §1 with $\gamma = 0.5$

**3.2 Case $M_1 \neq 0$**

Let us consider the case where value $M_1$ in the function of the magnetic permeability (1) of §2 is also nonzero. The equation for function $T(t)$ will be the following:

$$T'' + \frac{1 + 2M_0 t + 3M_1 t^2}{t\left(1 + M_0 t - M_1 t^2\right)} T' - \frac{\gamma^2 \left(1 + M_1 t^2\right)}{t^2 \left(1 + M_0 t - M_1 t^2\right)} T = 0. \tag{23}$$

Note that the solution of equation (23) shall transform into (19) at $M_1 = 0$. We search for the solution of (23) in the form of:

$$T^{(M_0, M_1)}(t) = t^\nu \sum_{k=0}^{+\infty} b_k \left(M_0, M_1, \nu\right) t^k. \tag{24}$$

Generating differentiation (24) and substituting the result into (23), we obtain:

$$\sum_{k=0}^{+\infty} t^{k+\nu} \Big\{ \nu(\nu-1) b_k t^k + \nu(\nu-1) M_0 b_k t^{k+1} + 2\nu b_{k+1}(k+1) t^{k+1} + b_{k+2}(k+2)(k+1) t^{k+2} + $$
$$+ 2M_0 \nu b_{k+1}(k+1) t^{k+2} + M_0 b_{k+2}(k+2)(k+1) t^{k+3} - M_1 \nu(\nu-1) b_k t^{k+2} -$$
$$- 2M_1 \nu b_{k+1}(k+1) t^{k+3} - M_1 b_{k+2}(k+2)(k+1) t^{k+4} + \nu b_k t^k + b_{k+1}(k+1) t^{k+1} + \tag{25}$$
$$+ 2M_0 \nu b_k t^{k+1} + 2M_0 b_{k+1}(k+1) t^{k+2} - 3M_1 \nu b_k t^{k+2} - 3M_1 b_{k+1}(k+1) t^{k+3} - \gamma^2 b_k t^k -$$
$$- \gamma^2 M_1 b_k t^{k+2} \Big\} = 0.$$

Let us transform the expression in the curly brackets (24), and we obtain:

$$\sum_{k=0}^{+\infty} t^{k+\nu} \left\{ b_k \left( \nu^2 - \gamma^2 \right) + t \left[ b_k \nu (\nu - 1) M_0 + 2\nu (k+1) b_{k+1} + (k+1) b_{k+1} + 2 M_0 \nu b_k \right] + \right.$$
$$+ t^2 \left[ b_{k+2} (k+2)(k+1) + 2 M_0 \nu (k+1) b_{k+1} - M_1 \nu (\nu - 1) b_k + 2 M_0 (k+1) b_{k+1} - \right.$$
$$\left. - 3 M_1 \nu b_k - \gamma^2 M_1 b_k \right] + t^3 \left[ M_0 b_{k+2} (k+2)(k+1) - 2 M_1 \nu b_{k+1} (k+1) - 3 M_1 (k+1) b_{k+1} \right] -$$
$$\left. - t^4 M_1 (k+2)(k+1) b_{k+2} \right\} = 0. \qquad (26)$$

Equating the coefficients at $t^\nu$ in expression (26), we obtain:

$$b_0 \left( \nu^2 - \gamma^2 \right) = 0, \; b_0 \neq 0 \Rightarrow \nu = \pm |\gamma|. \qquad (27)$$

Performing the same operation for coefficients $t^{\nu+1}$, $t^{\nu+2}$ taking (27) into account:

$$t^{\nu+1}: b_0 \nu (\nu - 1) M_0 + 2\nu b_1 + b_1 + 2 M_0 \nu b_0 = 0,$$
$$b_0 M_0 \left[ \nu (\nu - 1) + 2\nu \right] + b_1 (2\nu + 1) = 0,$$
$$b_1 = -b_0 M_0 \frac{\nu (\nu + 1)}{2\nu + 1}, \qquad (28)$$

$$t^{\nu+2}: b_1 M_0 \left[ \nu (\nu - 1) + 4\nu + 2 \right] - b_0 M_1 \left[ \nu (\nu - 1) + 3\nu + \gamma^2 \right] + b_2 4 (\nu + 1) = 0,$$
$$b_2 = \frac{-b_1 M_0 (\nu + 1)(\nu + 2) + b_0 2 M_1 \nu (\nu + 1)}{4 (\nu + 1)},$$
$$b_2 = \frac{-b_1 M_0 (\nu + 2) + b_0 2 M_1 \nu}{4}. \qquad (29)$$

Substituting the expression for coefficient $b_1$ (28) into (29), we obtain:

$$b_2 = b_0 M_0^2 \frac{\nu (\nu + 1)(\nu + 2)}{4 (2\nu + 1)} + \frac{b_0 M_1 \nu}{2} = b_0 \nu \frac{M_0^2 (\nu + 1)(\nu + 2) + 2 M_1 (2\nu + 1)}{4 (2\nu + 1)}. \qquad (30)$$

From (26) we obtain the recurrent formula for a random number $k$:

$$b_{k+4} \left( \nu^2 - \gamma^2 \right) + b_{k+3} \nu (\nu - 1) M_0 + 2\nu (k+4) b_{k+4} + (k+4) b_{k+4} + 2 M_0 \nu b_{k+3} +$$
$$+ b_{k+4} (k+4)(k+3) + 2 M_0 \nu (k+3) b_{k+3} - M_1 \nu (\nu - 1) b_{k+2} + 2 M_0 (k+3) b_{k+3} -$$
$$- 3 M_1 \nu b_{k+2} - \gamma^2 M_1 b_{k+2} + M_0 b_{k+3} (k+3)(k+2) - 2 M_1 \nu b_{k+2} (k+2) - 3 M_1 (k+2) b_{k+2} -$$
$$- M_1 (k+2)(k+1) b_{k+2} = 0,$$

or

$$b_{k+3} M_0 \left[ \nu (\nu - 1) + 2\nu + 2\nu (k+3) + 2 (k+3) + (k+3)(k+2) \right] +$$
$$+ b_{k+4} \left[ 2\nu (k+4) + (k+4) + (k+4)(k+3) \right] +$$
$$b_{k+2} M_1 \left[ -\nu (\nu - 1) - 3\nu - \gamma^2 - 2\nu (k+2) - 3 (k+2) - (k+2)(k+1) \right] = 0,$$

or replacing $k \to k-2$, we obtain:

$$b_{k+2} = \frac{-b_{k+1}M_0(v+k+1)(v+k+2)}{(k+2)(2v+k+2)} + \frac{b_k M_1\left[(v+k+1)^2 + v^2 - 1\right]}{(k+2)(2v+k+2)}. \tag{31}$$

By recurrent formula (31) using formulas (28) and (30) we can obtain all the coefficients of the series, thereby to determine the solution of equation (23). The coefficients of (31) depend on parameters $M_0$ and $M_1$, unlike the coefficients of (18) which depend only on $M_0$. Note that when $M_1 = 0$, the coefficients of (31) become the coefficients of (18):

$$b_k(M_0, M_1, v)\big|_{M_1=0} = a_k(M_0, v).$$

Fig.10 shows the comparison of graphs of solution (24) and solution (19) for values $\gamma = 3$, $M_0 = 0.1$, $M_1 = 0.01$.

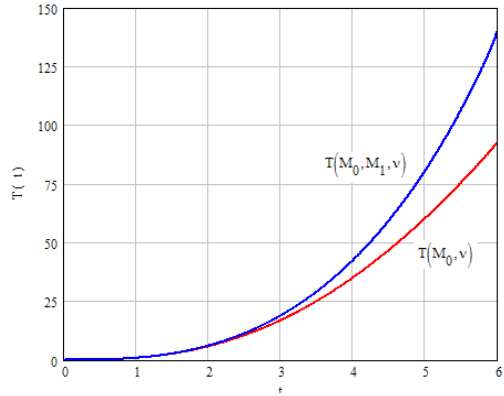

Figure 10. Comparison solution (24) and solution (19)

### §4 Boundary value problem for non-linear elliptic equation

Let us consider a boundary value problem (1) in a domain with corner (Fig. 11).

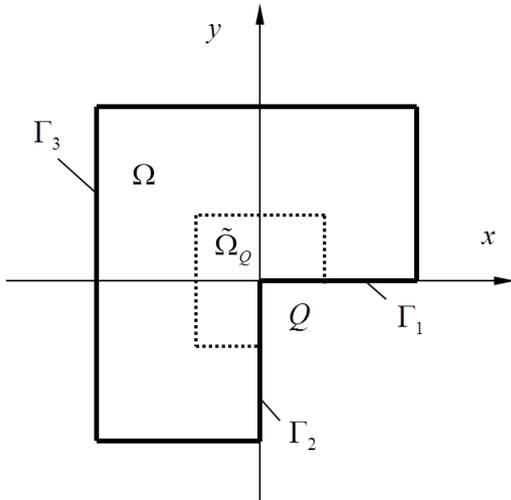

Figure 11. Corner domain

$$\text{div}\left[\mu(|\nabla u|)\nabla u\right] = 0, \quad p \in \Omega \tag{1}$$

$$u\big|_\Gamma = \Psi(p), \quad p \in \Gamma \tag{2}$$

$$\Psi(p) = \begin{cases} 0, & p \in \Gamma_1 \\ g(p), & p \in \Gamma_2 \end{cases}, \tag{3}$$

where $\Gamma = \Gamma_1 \cup \Gamma_2 \cup \Gamma_3$, and $\Psi \in C^{(1)}(\Gamma)$. The function $\mu$ in equation (1) meets requirements (1) from §2 and $a(r) > 0$, which renders equation (1) to be elliptic. Let $u(p)$ be a solution of (1-3). We are going to assess $|\nabla u|$ around point $Q$ (coordinate origin).

We assume that the solution $u(p)$ meets the following requirements:

$\exists \delta > 0:$

$$\forall p \in S_\delta(Q) \cap \Omega: \ J(p) = u_{xx}(p)u_{yy}(p) - u_{xy}^2(p) \neq 0, \quad (4)$$
$$\forall p_1, p_2 \in S_\delta(Q) \cap (\Omega \cup \Gamma), \ p_1 \neq p_2: \ \nabla u(p_1) \neq \nabla u(p_2).$$
$$u_x(0, y), u_y(x, 0) \text{ are monotonous functions if } x \in \Gamma_1, y \in \Gamma_2, |x| < \delta, |y| < \delta.$$

It should be noted that such solutions exist, it can be seen from §3.

### 4.1 Assessment of $|\nabla u|$ for boundary value problem

Applying the Legendre transformation for function $u(p)$ under $p \in S_\delta(Q) \cap (\Omega \cup \Gamma)$ «boundary $\Gamma_1$» $(\omega_\eta = 0, \omega_\xi > 0, \xi = 0)$ will be transformed in «boundary $\tilde{\Gamma}_1$», and «boundary $\Gamma_2$» $(\omega_\xi = 0, \omega_\eta < 0, \eta = 0)$ in «boundary $\tilde{\Gamma}_2$». The signs of $\eta = u_y(x, 0)$ for «boundary $\Gamma_1$» and $\xi = u_x(0, y)$ for «boundary $\Gamma_2$» are unclear. Now we consider the case when $\eta < 0$ and $\xi > 0$ for boundaries $\tilde{\Gamma}_1$ and $\tilde{\Gamma}_2$ respectively. Applying the transformation we will get domain $\tilde{\Omega}_1$ or $\tilde{\Omega}_2$, illustrated in Fig. 12. Firstly, let us consider domain $\tilde{\Omega}_2$. Substituting variables $r = 1/t$ (see §3), we will come to the following boundary value problem (Fig. 13)

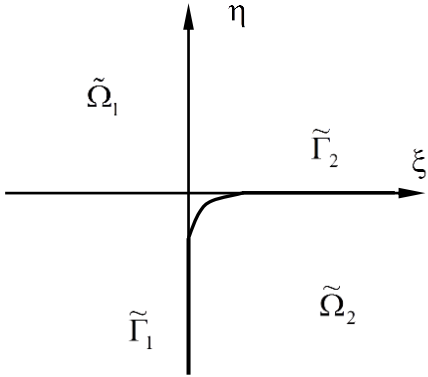 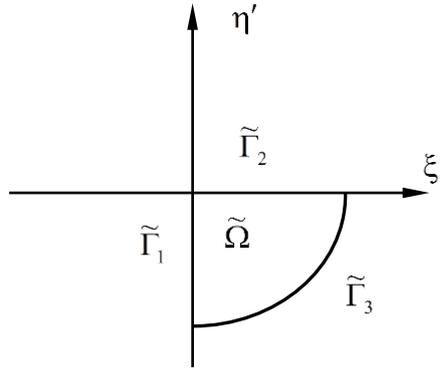

Figure 12. Domains $\tilde{\Omega}_1$ and $\tilde{\Omega}_2$     Figure 13. Domain $\tilde{\Omega}$

$$L_{\bar{a}} w(p) = 0, \quad p \in \tilde{\Omega},$$
$$w|_{\tilde{\Gamma}} = \tilde{\Psi}(p), \quad p \in \tilde{\Gamma}, \quad (5)$$

where

$$L_{\bar{a}} = \frac{\partial^2}{\partial t^2} + \frac{1}{t}(2 - \bar{a}(t))\frac{\partial}{\partial t} + \frac{\bar{a}(t)}{t^2}\frac{\partial^2}{\partial \varphi^2}, \quad \bar{a}(t) = a\left(\frac{1}{r}\right). \quad (6)$$

Since $\omega_\eta = 0$ on $\tilde{\Gamma}_1$ then $\omega|_{\tilde{\Gamma}_1} = 0$ and similarly $\omega|_{\tilde{\Gamma}_2} = 0$. Thus we get

$$\tilde{\Psi}(p) = \begin{cases} 0, & p \in \tilde{\Gamma}_1 \cup \tilde{\Gamma}_2 \\ \tilde{g}(p), & p \in \tilde{\Gamma}_3 \end{cases} \quad (7)$$

where $\tilde{\Gamma} = \tilde{\Gamma}_1 \cup \tilde{\Gamma}_2 \cup \tilde{\Gamma}_3$, and $\tilde{\Psi} \in C^{(1)}(\tilde{\Gamma})$. Function $\tilde{\Psi}$ is defined by $u(p)$ from the Legendre transformation. Now we will consider the solutions of (5-7), possessing the limited gradients. Doing (5-7) by the factorization method $w(p) = T(t)\Phi(\varphi)$, we come to

$$T'' + \frac{2-\bar{a}(t)}{t}T' - \lambda^2 \frac{\bar{a}(t)}{t^2}T = 0, \quad T(0) = 0, \tag{8}$$

$$\Phi'' + \lambda^2 \Phi = 0, \quad \Phi\left(\frac{3\pi}{2}\right) = \Phi(2\pi) = 0.$$

It results in

$$\Phi_k(\varphi) = \sin\left(\lambda_k\left(\varphi - \frac{3\pi}{2}\right)\right), \quad \lambda_k = 2k, \quad k = 1,\ldots \tag{9}$$

$$T_k^{(1)}(t) = t^{|\lambda_k|}\sum_{s=0}^{+\infty} a_s(\lambda_k) t^s, \tag{10}$$

$$T_k^{(2)}(t) = A T_k^{(1)}(t) \ln t + t^{-|\lambda_k|}\sum_{s=0}^{+\infty} d_s(\lambda_k) t^s.$$

Functions $T_k^{(2)}(t)$ will be absent when decomposing in a series of (5-7). Consequently, we get:

$$\bar{w}(t,\varphi) = \sum_k \bar{C}_k T_k^{(1)}(t)\Phi_k(\varphi). \tag{11}$$

Thus, for function $v(r,\varphi)$ under $r > r_0$ (where $1/r_0 = t_0$) we get:

$$v(r,\varphi) = \sum_{k=1}^{+\infty} \bar{C}_k T_k\left(\frac{1}{r}\right)\Phi_k(\varphi). \tag{12}$$

Now let us assess $|\nabla u|$. In order to do it we will assume $v$ under $r > r_0$ as:

$$v(r,\varphi) = \Phi_1(\varphi)\frac{\bar{C}_1}{r^{|\lambda_1|}} a_0(\lambda_1) + O\left(\frac{1}{r^{|\lambda_1|+1}}\right). \tag{13}$$

then we can derive:

$$\lim_{r\to\infty} r^{|\lambda_1|+1}|\nabla v| = |\lambda_1||\bar{C}_1 a_0(\lambda_1)| = Const. \tag{14}$$

Since $r = |\nabla u|$ and $\bar{r} = \sqrt{x^2 + y^2} = |\nabla v| \equiv |\nabla \omega|$, then

$$|\nabla u| \sim \bar{r}^{-\frac{1}{1+|\lambda_1|}} \text{ under } \bar{r} \to 0. \tag{15}$$

Since $\lambda_1 = 2$ then the rate of the gradient increase for (1-3) around the corner point is $\bar{r}^{-1/3}$, i.e., the same as that for the similar boundary value problem with the Laplace equation [10, 18]. In case of considering domain $\tilde{\Omega}_1$ instead of domain $\tilde{\Omega}_2$ (Fig. 11), the transformation loses its uniqueness. The case when $\eta > 0$ and $\xi < 0$ for boundaries $\tilde{\Gamma}_1$ and $\tilde{\Gamma}_2$ respectively is similar to the previous one and the same assessment on $|\nabla u|$ is valid. In other cases when:

- $\eta < 0$ for $\Gamma_1$, $\xi < 0$ for $\Gamma_2$;
- $\eta > 0$ for $\Gamma_1$, $\xi > 0$ for $\Gamma_2$;

the transformation also will lose its uniqueness.

### 4.2. Numerical algorithm for boundary value problem

Now a numerical method for the solution of the boundary value problem (1-3) will be described. An algorithm for the solution of the problem will be designed in the similar way that is described in [45] for the Laplace equation. A difference scheme will be built only for the neighborhood of the corner point $\tilde{\Omega}_Q$

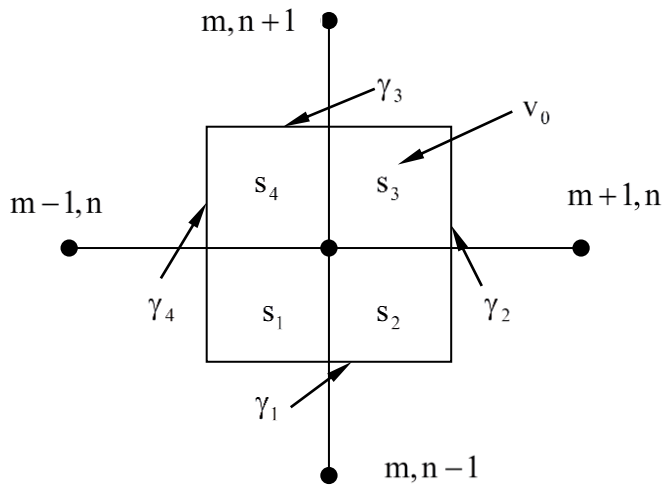

Figure 14. Difference scheme

Integration of equation (1) by volume $v_0$ (Fig. 14) yields

$$\int_{v_0} div\left[\mu(|\nabla u|)\nabla u\right]dxdy = \oint_\gamma \mu\frac{\partial u}{\partial y}dx - \mu\frac{\partial u}{\partial x}dy = 0, \quad (16)$$

where $\gamma$ is the boundary of domain $v_0$ and $\gamma = \bigcup_{m=1}^{4}\gamma_m$. Only approximation of $\int_{\gamma_1}\mu\frac{\partial u}{\partial y}dx$ will be considered, approximation for the other contours is similar. Function $P(r,\varphi)$ is assumed to be derived from function $T_1^{(1)}\left(\frac{1}{r}\right)\Phi_1(\varphi)$ applying the Legendre transformation. Let $G_\alpha = \frac{\partial P}{\partial x_\alpha}$ be for $\alpha = 1,2$. From the above, function $G_\alpha$ contributes the main part in $|\nabla u|$ of (1-3) around the corner point, therefore we will apply the following approximation

$$\frac{\partial u(p)}{\partial x_\alpha} \approx C_1 G_\alpha(p), \quad \alpha = 1,2, \quad p \in \tilde{\Omega}_Q, \quad (17)$$

where $C_1 = Const$ is to be determined. Mean values for $G_\alpha$ are assumed as

$$G_\alpha^\beta = \frac{1}{\Delta s_\beta}\int_{\Delta s_\beta} G_\alpha dx_\beta, \quad \alpha = 1,2 \quad \beta = 1,2, \quad (18)$$

where $\Delta s_\beta$ is the segment parallel to X- or Y- axes for $\beta = 1$ or $\beta = 2$ respectively. Let us assume function $\mu(|\nabla u|)$ to be piecewise constant in segments $s_i$, $i = 1\ldots 4$ (Fig. 14) and to be equal to $\mu_i$, $i = 1\ldots 4$ accordingly.

$$\int_{\gamma_1} \mu \frac{\partial u}{\partial y} dx \approx \mu_1 \int_{x_m - h_m^x/2}^{x_m} \frac{\partial u}{\partial y} dx + \mu_2 \int_{x_m}^{x_m + h_{m+1}^x/2} \frac{\partial u}{\partial y} dx. \qquad (19)$$

We consider only the first integral term, the second one is considered in the similar way.

$$\int_{x_m - h_m^x/2}^{x_m} \frac{\partial u}{\partial y} dx \approx C_1 \int_{x_m - h_m^x/2}^{x_m} G_2 dx = \Delta s_1 G_2^1 C_1, \quad \Delta s_1 = \frac{h_m^x}{2},$$

$$u(x_m, y_n) - u(x_m, y_{n-1}) = \int_{y_{n-1}}^{y_n} \frac{\partial u}{\partial y}(x_m, y) dy \approx C_1 \int_{y_{n-1}}^{y_n} G_2 dy = C_1 \Delta s_2 G_2^2, \qquad (20)$$

where $\Delta s_2 = h_n^y$, thus

$$C_1 \approx \frac{1}{G_2^2} \frac{u_{m,n} - u_{m,n-1}}{h_n^y}. \qquad (21)$$

As a result we get

$$\mu_1 \int_{x_m - h_m^x/2}^{x_m} \frac{\partial u}{\partial y} dx \approx \mu_1 \frac{G_2^1}{G_2^2} \frac{h_m^x}{2} \frac{u_{m,n} - u_{m,n-1}}{h_n^y} = g_1 (u_{m,n} - u_{m,n-1}),$$

$$\mu_2 \int_{x_m}^{x_m + h_{m+1}^x/2} \frac{\partial u}{\partial y} dx \approx g_2 (u_{m,n} - u_{m,n-1}),$$

$$\int_{\gamma_1} \mu \frac{\partial u}{\partial y} dx \approx g_x^{(-)} (u_{m,n} - u_{m,n-1}), \text{ where } g_x^{(-)} = g_1 + g_2. \qquad (22)$$

Similar expressions can be derived for the other contours. Consequently, (16) takes the following form:

$$u_{m,n} = \frac{g_x^{(-)} u_{m,n-1} + g_x^{(+)} u_{m,n+1} + g_y^{(+)} u_{m+1,n} + g_y^{(-)} u_{m-1,n}}{g_x^{(-)} + g_x^{(+)} + g_y^{(-)} + g_y^{(+)}}. \qquad (23)$$

Now we present a boundary value problem solved by the similar way. Function $\mu$ was assumed as (1) from §2 and function $\bar{w}(t, \varphi) = T_1^{(1)}(t)\Phi_1(\varphi) + T_2^{(1)}(t)\Phi_2(\varphi)$. Comparative results of calculation of relative errors are presented below. These results were obtained with the use of the above algorithm (Fig. 15) (23) and without it (Fig. 16) (24).

$$g_1 = \mu_1 \frac{h_m^x}{2h_n^y}, \quad g_2 = \mu_2 \frac{h_{m+1}^x}{2h_n^y}, \quad g_x^{(-)} = g_1 + g_2 = \frac{1}{2h_n^y}(\mu_1 h_m^x + \mu_2 h_{m+1}^x). \qquad (24)$$

As it can be seen from the figures, the relative errors around the corner point decrease by about 10 times.

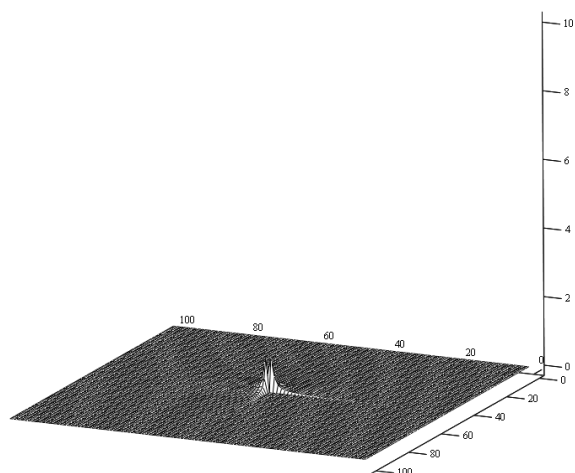 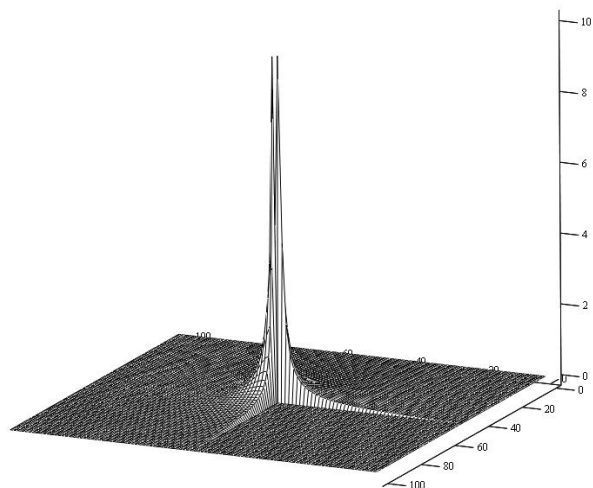

Figure 15. Relative error for (23), (22)    Figure 16. Relative error for (23), (24)